\documentclass[aps,prb,twocolumn,superscriptaddress,showpacs]{revtex4-1}

\usepackage{graphicx}
\usepackage{calc}
\usepackage{bm}
\usepackage{color}
\usepackage{textcomp}

\begin{document}

\title{Josephson spin-valve realization in the magnetic nodal-line topological semimetal Fe$_3$GeTe$_2$}

\author{O.O.~Shvetsov}
\author{Yu.S.~Barash}
\author{A.V.~Timonina}
\author{N.N.~Kolesnikov}
\author{E.V.~Deviatov}
\affiliation{Institute of Solid State Physics of the Russian Academy of Sciences, Chernogolovka, Moscow District, 2 Academician Ossipyan str., 142432 Russia}

\date{\today}

\begin{abstract}
    Three-dimensional  van der Waals ferromagnet Fe$_3$GeTe$_2$ (FGT) is regarded as a candidate for the magnetic topological nodal line semimetal. We investigate lateral electron transport between two 3~$\mu$m spaced superconducting In leads beneath a thick three-dimensional FGT exfoliated flake. At low 30~mK temperature, we observe Josephson supercurrent that exhibits unusual critical current $I_c$ suppression by the magnetic field $B$. The overall $I_c(B)$ pattern is asymmetric in respect to the B sign. We demonstrate, that the asymmetry is defined by the magnetic field sweep direction, so the $I_c(B)$ pattern is strictly reversed (as $B$ to $-B$ inversion) for the opposite sweeps. We also observe an interplay between maximum and minimum in $I_c(B)$ in normal magnetic fields, while there are fast aperiodic $I_c(B)$ fluctuations for the in-plane ones. These effects can not be expected for homogeneous superconductor-ferromagnet-superconductor junctions, while they are known for Josephson spin valves. The mostly possible scenario for Josephson spin valve realization in FGT is the misalignment of spin polarizations of the Fermi arc surface states and ferromagnetic FGT bulk, but we also discuss possible influence of spin-dependent transport between  magnetic domains.
    
\end{abstract}

\pacs{73.40.Qv  71.30.+h}

\maketitle

\section{Introduction}

Recently, Fe$_3$GeTe$_2$ (FGT) has attracted significant attention as a promising platform for  novel physical phenomena, which are connected with magnetic and electronic non-trivial topology. FGT is an itinerant van der Waals ferromagnet characterized by an out-of-plane magnetocrystalline anisotropy both for three-dimensional single crystals and down to two-dimensional limit, which was confirmed by theoretical and experimental investigations~\cite{2dFM1,2dFM2,2dFM3,anisotrop,kerr,fgttc}. Experimentally, FGT shows large anomalous Hall~\cite{ahe1,kim} and Nernst~\cite{Nernst} effects, topological Hall effect~\cite{PTHE} and Kondo lattice physics~\cite{kondo}. From the view of the electronic band structure, three-dimensional FGT is a unique candidate for the ferromagnetic nodal line semimetal~\cite{kim}, hosting spin-polarized Fermi arc surface states~\cite{asymmr}.

Different realizations of spin valves are known for magnetic materials.  Usually, spin valves are realized as ferromagnetic multilayers~\cite{valve1,valve2} with different layers' thickness. The multilayer resistance depends on the  mutual orientation of their magnetizations due to the spin-dependent scattering, so the resistance can be affected by external magnetic field or high current density. Due to the different spin polarization of the Fermi arc surface states and ferromagnetic bulk, magnetic topological materials should also demonstrate spin-valve transport properties~\cite{sot_review,molenkamp2021}, i.e. they can be regarded as natural realization of spin-valves. In this case, spin-polarized Fermi arcs and ferromagnetic bulk represent thin (free) and thick (reference) layers, respectively~\cite{timnal,wte,cosns,bite}. 

In proximity with a superconductor, topological surface (or edge) states are able to carry supercurrents over extremely large distances~\cite{topojj1,topojj2,topojj3,topojj4,topojj5}. For the magnetic topological materials it naturally implies spin triplet superconductivity, which is the mutual effect of superconductivity, exchange interaction and  spin-orbit coupling~\cite{cro2,triplet-theor,review-triplet,dutta,spin-orbit-triplet1,spin-orbit-triplet2}. Triplet supercurrent can be expected, e.g., for a Josephson spin valve~\cite{krasnov,reverse,jsv1,jsv2,jsv3} (JSV), where ferromagnetic multilayer is sandwiched between two superconducting electrodes. In the majority of devices the Josephson current is directed perpendicular to the layers, but the spin-valve effects and, in particular, the generation of the triplet supercurrent can also occur in systems, where the supecurent flows along the planes \cite{golubov}.  

In JSVs supercurrent is defined mainly by the relative orientation of the layers' magnetizations, while in conventional Josephson junctions it is modulated by magnetic flux.  The strength of the singlet-triplet conversion substantially depends on the particular configuration of the magnetization misalignment in the Josephson spin valve. For the supercurrent flowing perpendicular to the layers, such a dependence on relative orientations of the layers' magnetizations was studied in detail and has been recently used for experimental identification of relative weights of singlet and triplet amplitudes constituting a net supercurrent \cite{krasnov}.   Due to the natural spin-valve realization, magnetic topological semimetals like FGT may be regarded as a platform for planar JSV investigations.

Symmetry analysis and first principles calculations have shown, that the inversion symmetry breaking can occur at the FGT interface~\cite{skyrm4}.
Noncentrosymmetric interfacial effects are known to be able to substantially influence the charge transport in magnetic systems, in particular via the spin-orbit torque, and to result in unidirectional transport properties~\cite{gambardella2015,qiu2018,gambardella2019,kurebayashi2014,molenkamp2021}. In proximity with
superconductivity, broken inversion and time reversal symmetry can generaly lead to asymmetries of the Josephson current with respect to the magnetic field reversal, e.g., due to chiral properties of the topologically protected states~\cite{chenetal2018,fu2021}. In superconducting heterostructures with 
noncomplanar magnetization textures, breaking the magnetization reversal symmetry can result in the direct coupling between the magnetic moment and the supercurrent, and in the anomalous Josephson effect~\cite{buzdin2008,buzdin2009,linder2014,bergeret2017,buzdin2017}.

Here, we investigate lateral electron transport between two 3~$\mu$m spaced superconducting In leads beneath a thick three-dimensional FGT exfoliated flake. At low 30~mK temperature, we observe Josephson supercurrent that exhibits unusual critical current $I_c$ suppression by the magnetic field $B$. The overall $I_c(B)$ pattern is asymmetric in respect to the B sign. We demonstrate, that the asymmetry is defined by the magnetic field sweep direction, so the $I_c(B)$ pattern is strictly reversed (as $B$ to $-B$ inversion) for the opposite sweeps. We also observe an interplay between maximum and minimum in $I_c(B)$ in normal magnetic fields, while there are fast aperiodic $I_c(B)$ fluctuations for the in-plane ones. These effects can not be expected for homogeneous superconductor-ferromagnet-superconductor junctions, while they are known for Josephson spin valves. The mostly possible scenario for Josephson spin valve realization in FGT is the misalignment of spin polarizations of the Fermi arc surface states and ferromagnetic FGT bulk, but we also discuss possible influence of spin-dependent transport between  magnetic domains.

\section{Samples and technique}

\begin{figure}
\includegraphics[width=\columnwidth]{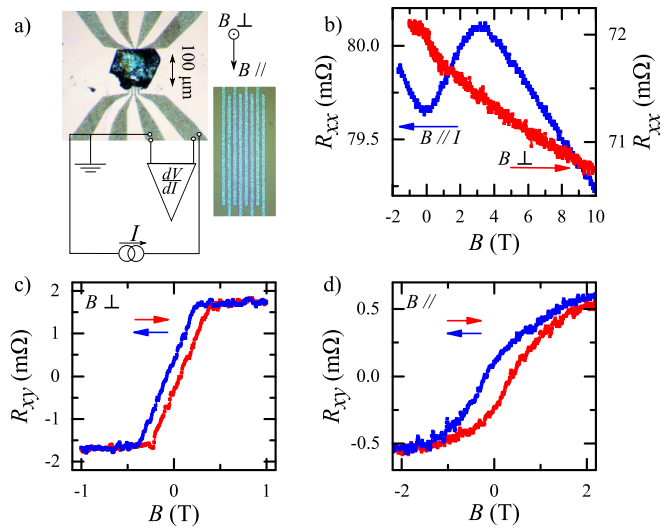}
\caption{(Color online)  (a) A top-view image of the sample with electrical connections. A thick (1~$\mu$m) single-crystal FGT flake is placed by the flat bottom surface on the pre-defined superconducting In leads. The right inset shows the initial leads pattern, which consists of 10~$\mu$m wide indium stripes separated by 3~$\mu$m intervals. Electron transport is investigated between two neighbor In leads in a standard four-point technique, all the wire resistances are excluded. Arrows indicate the in-plane $B_{||}$ and normal $B_{\perp}$ magnetic field orientations for Figs.~\ref{3} and ~\ref{4}. 
(b,c,d) Magnetoresistance measurements, to confirm FGT quality, for a reference sample with Au leads in standard Hall bar geometry. 
(b) Longitudinal magnetoresistance $R_{xx}$ for the in-plane (left axis, blue curve) and for the normal (right axis, red curve) fields.  (c,d) Hall $R_{xy}(B)$  hysteresis loops in normal and in-plane fields, respectively, which is usually ascribed to anomalous and topological Hall effects in FGT~\cite{PTHE}. The arrows denote magnetic field sweep directions.}
\label{infgt_sample}
\end{figure}

Fe$_3$GeTe$_2$ was synthesized from elements in evacuated silica ampule in a two-step process. At the first step, the load was heated up to 470$^\circ$~C  at 10 deg/h rate and the ampule was held at this temperature for 50 h. At the second step, the temperature was increased up to 970$^\circ$~C with the same rate. After 140 h exposure, the ampule was cooled down to the room temperature at 5 deg/h rate. X-ray diffraction data indicates, that the iron tellurides FeTe and FeTe2  were also found in the obtained material, in addition to the expected Fe$_3$GeTe$_2$ compound. 

To obtain Fe$_3$GeTe$_2$ single crystals, the synthesized mixture  was sealed in evacuated silica ampule with some admixture of iodine. The transport reaction was carried out for 240 h with temperatures  530$^\circ$~C and  410$^\circ$~C in hot and cold zones, respectively. Afterward,  the  ampule was quenched in a liquid nitrogen. Water-solvable iron and tellurium iodides were removed in hot distilled water from the obtained Fe$_3$GeTe$_2$ single crystals, so the X-ray diffraction analysis confirmed strict Fe$_3$GeTe$_2$ composition.   

Non-trivial surface properties are known for three-dimensional topological semimetal single crystals~\cite{armitage}. Thus, we use thick (1~$\mu$m) FGT flakes, which are obtained by a mechanical cleavage from the initial single crystal. 

Fig.~\ref{infgt_sample}(a) shows a top-view image of a FGT flake with underlying  indium leads. The leads pattern is formed by lift-off technique after thermal evaporation of 100~nm In   on the insulating SiO$_2$ substrate. The 10~$\mu$m wide  leads are separated by 3~$\mu$m intervals.  One FGT flake is transferred to the substrate with the defined In  leads pattern and pressed to the leads slightly. No stress is needed for a flake to hold on the In leads afterward. This procedure allows to create  transparent FGT-In interfaces~\cite{cosnsjc,cdas,wte2} without mechanical polishing or chemical treatment, and to protect the relevant (bottom) FGT surface from any oxidation or contamination.
 
To confirm FGT quality, magnetoresistance measurements are performed also in standard Hall bar geometry for reference samples with normal (Au) leads.   In Fig.~\ref{infgt_sample}(b),  longitudinal magnetoresistance $R_{xx}$ is monotonous and negative in normal magnetic fields (red curve, right axis), while it shows a kink at 3.5~T for the in-plane configuration (blue curve, left axis). This behavior coincides well with the previously reported results~\cite{PTHE}. 
Moreover, large anomalous Hall effect is shown in Fig.~\ref{infgt_sample}(c) for normal field orientation, while hysteresis in $R_{xy}$ is also known for the in-plane field as a novel planar Hall effect, see   Fig.~\ref{infgt_sample}(d). The latter has been also recognized as topological Hall effect related to the complicated spin structures in FGT~\cite{PTHE}. 

We study electron transport between two neighbor In leads in a standard four-point technique, see Fig.~\ref{infgt_sample}(a). All the wire resistances are excluded, which is necessary for low-impedance samples. To obtain $dV/dI(I)$ characteristics, dc current is additionally modulated by a low 2~$\mu$A (below the dc current step) ac component at a 1107~Hz frequency. We measure the  ac component of the potential drop ($\sim dV/dI$) by lock-in. The signal is confirmed to be independent of the modulation frequency within 100 Hz -- 10kHz range, which is defined by the applied filters.  The measurements are performed  within the 30~mK -- 1.2~K temperature range.

\section{Experimental results}

\begin{figure}
\includegraphics[width=\columnwidth]{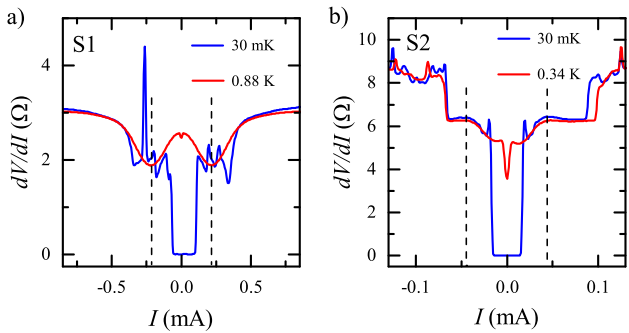}
\caption{(Color online)  Josephson effect for   two different samples with  In-FGT-In junctions (S1 and S2 in (a) and (b), respectively). Qualitative behavior is similar, despite strongly different  critical current ($I_c$=0.17~mA in (a)  and 0.018~mA in (b)) and normal resistance values.  The zero-resistance state appears below  0.88~K for S1 and 0.34~K for S2.  The high-temperature curves are typical for Andreev reflection. The superconducting gap  positions are denoted by the dashed lines (see the main text), it should not be confused with asymmetric jumps in $dV/dI$ at much higher currents. The data are presented for zero magnetic field.
 }
\label{2}
\end{figure}

\begin{figure}
\includegraphics[width=\columnwidth]{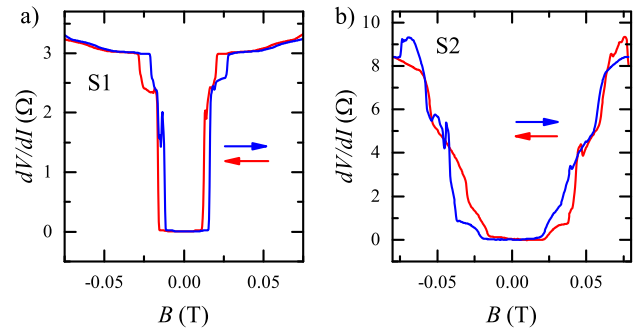}
\caption{(Color online) Influence of the external in-plane (a) and normal (b) magnetic fields on Josephson effect at $T$ = 30~mK for S1 and S2, respectively.
$dV/dI(B)$ curves are not symmetric with respect to zero field, the observed asymmetry depends on the magnetic field sweep direction, which  is denoted by arrows of the corresponding color. All the $dV/dI$ features are  mirrored for the opposite field sweeps, so $dV/dI(B)$ curves are strictly reversed for two sweep directions.  This curve reversal can not be expected for  a superconductor-ferromagnet-superconductor junction with the homogeneous magnetization of the ferromagnetic layer, but it is a fingerprint of the complicated spin structures. The data are obtained at 30~mK.}
\label{3}
\end{figure}

\begin{figure*}
    \includegraphics[width=500pt]{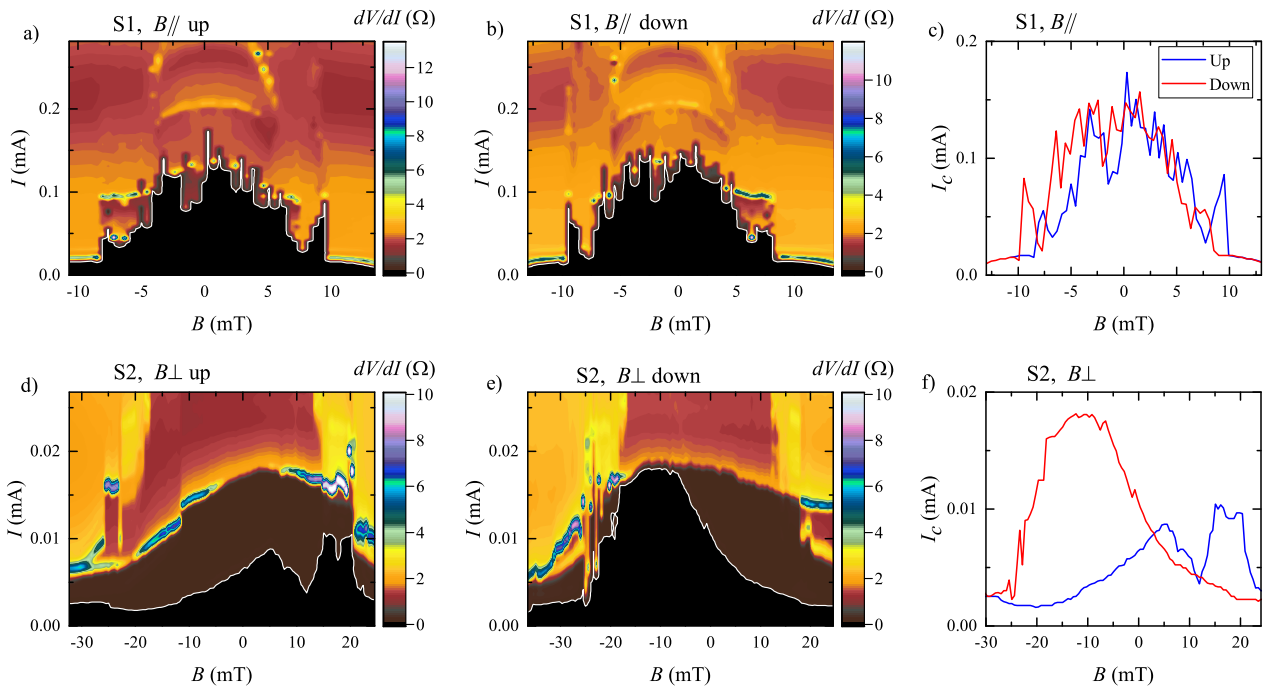}
   \caption{(Color online)  Colormaps of $dV/dI(I,B)$ for samples S1 and S2 in (a,b) and (d,e) respectively. The panels (a,d) and (b,e) differ by the magnetic field sweep direction, which is from negative to positive values in (a) and is just opposite in (b). All the data are obtained at 30~mK. To establish definite sample magnetization, every magnetic field sweep begins from high field value $B$ = $\pm$100~mT (the sign depends on the sweep direction). The colormaps are obtained from $dV/dI(I)$ curves at fixed magnetic field values, which are  changed point-by-point in up or down directions. To establish definite sample magnetization state, every magnetic field sweep cycle begins from high field value $B$ = $\pm$100~mT. The $dV/dI(B)$ reversal from Fig.~\ref{3} can be clearly seen, e.g., by the asymmetric black feature at $\pm$9~mT in (a,b). The reversal effect is even more pronounced in (d,e) for normal magnetic fields.   
   (c,f) $I_c(B)$ dependences for the in-plane and normal magnetic field orientations, respectively. The general $I_c(B)$ shapes are asymmetric in both cases,  the asymmetry is reversed for the up (blue) and down (red) field sweeps. For the in-plane magnetic fields, $I_c(B)$ shows well-reproducible aperiodic fluctuations in (c). On the contrary, no noticeable fluctuations can be observed  in (f).  In normal magnetic fields,  there is an interplay between maximum and minimum in $I_c(B)$ at $\pm$12~mT, which is well known for the Josephson spin valves~\cite{krasnov,reverse,jsv1,jsv2,jsv3}. The data are obtained at 30~mK.
   }
    \label{4}
\end{figure*}

\begin{figure}
\includegraphics[width=\columnwidth]{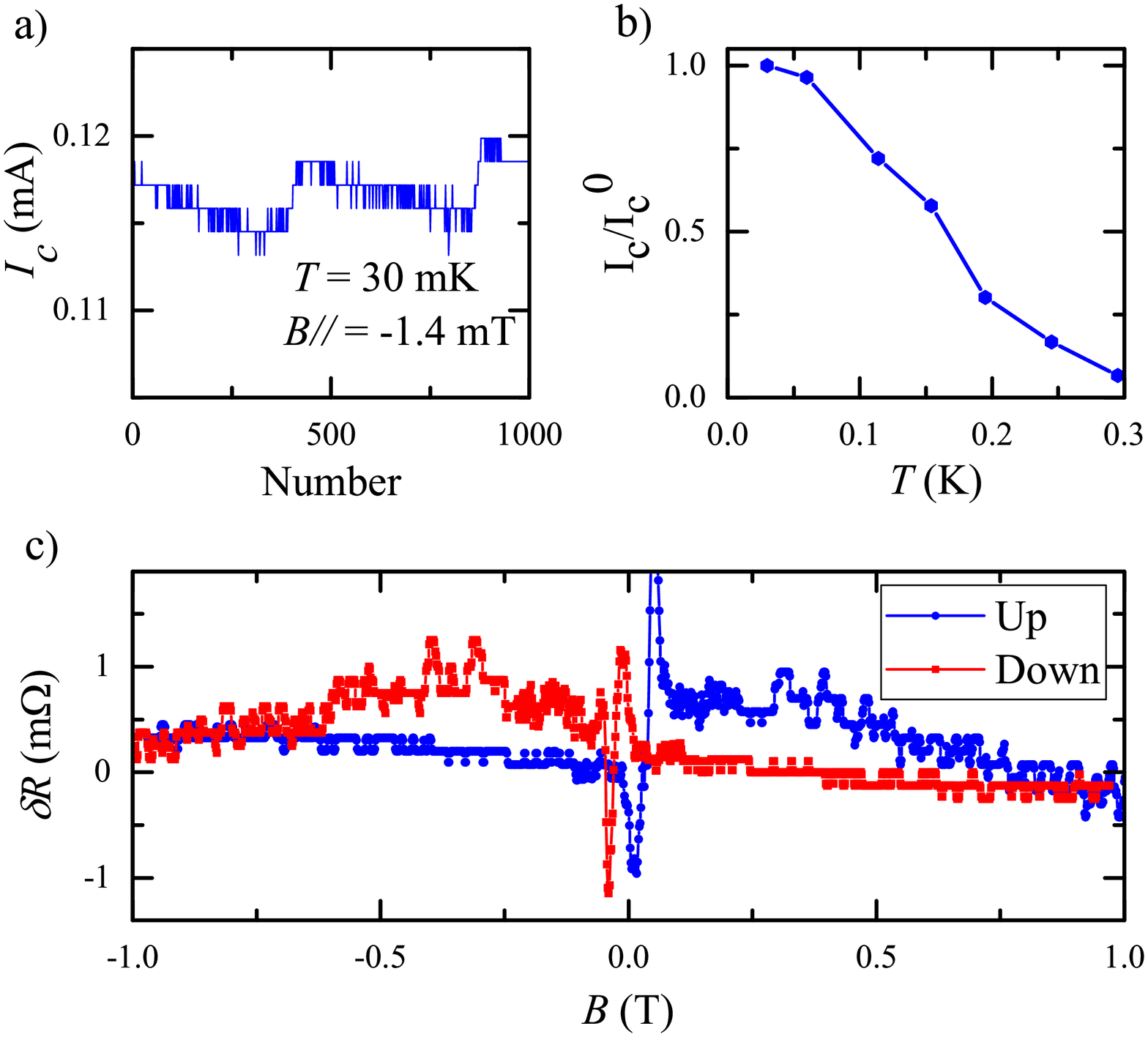}
\caption{(Color online) (a) Stability of $I_c$, as it is demonstrated for 1000 sequentially recorded curves at a fixed in-plane field value $B$ = -1.4~mT. The maximum $I_c$ deviation  is about 0.005~mA at $B$ = -1.4~mT, which is negligible comparing to the fluctuations' amplitude 0.05~mA in Fig.~\ref{4} (c). (b) Temperature dependence of $I_c$ for S2 in zero field, which  supports a large contribution of triplet supercurrent in our In-FGT-In junctions~\cite{cro2}. (c) Typical spin-valve hysteresis~\cite{valve1,valve2} in magnetoresistance of a single Au-FGT junction for the reference FGT flake. Blue and red curves correspond to the up and down magnetic field sweeps, respectively. In FGT, spin-polarized surface state  acts as a source of spin, while spin-dependent scattering within the ferromagnet results in the resistance dependence on the magnetization direction. 
 }
\label{5}
\end{figure}

Fig.~\ref{2}  clearly demonstrates Josephson effect for   two different  samples, which are referred as S1 and S2. Qualitative behavior is similar, despite strongly different  critical current $I_c$ and normal resistance values. 

As expected, the zero-resistance state appears below some critical temperature, which is about 0.88~K and 0.34~K for the devices in Fig.~\ref{2} (a) and (b). These In-FGT-In junctions are characterized by different maximum supercurrent values $I_c$=0.17~mA (S1) and 0.018~mA (S2). 
 
The high temperature curves are typical for Andreev reflection in Fig.~\ref{2}. The superconducting gap  positions are defined by symmetric resistive $dV/dI$ features at low currents, they are denoted by   dashed lines in Fig.~\ref{2}. For S1, $\Delta_{S1} = 0.42$~meV is obtained from $\pm$0.22~mA $dV/dI$ feature  positions and 1.9~$\Omega$ resistance level in Fig.~\ref{2} (a).  $\Delta_{S2}$ can also be estimated as  0.28~meV in Fig.~\ref{2} (b). These gap values are reasonable for In-FGT-In junctions, since the bulk 0.5~meV In  gap  should be partially suppressed by the intrinsic FGT magnetization.   
 
Since FGT is an uniaxial ferromagnet, which is confirmed by the Hall curves in Fig.~\ref{infgt_sample} (c,d), it seems to be reasonable to investigate Josephson effect in differently oriented magnetic fields. On the other hand, In-FGT junctions are known to be badly reproduced in different coolings, which restricts the possibilities to remount a sample in the dilution refrigerator. For these reasons, qualitatively similar samples S1 and S2 are initially mounted in the in-plane and normal field orientations, respectively, to avoid unwanted influence of the cooling procedure on the experimental data. 

Fig.~\ref{3} demonstrates the influence of external in-plane (a) and normal (b) magnetic fields on sample resistance at $T$ = 30~mK for S1 and S2, respectively. 
The result is qualitatively similar for both field orientations: the zero-resistance state is suppressed by the external field,  $dV/dI(B)$ curves are not symmetric with respect to the zero field value.  

As a most important, the observed $dV/dI(B)$ asymmetry depends on the magnetic field sweep direction. Moreover, all the $dV/dI$ features are  mirrored for the opposite (blue and rad colors) field sweeps, so $dV/dI(B)$ curves are strictly reversed for the two sweep directions in Figs.~\ref{3} (a) and (b).  This curve reversal can not be expected for  a superconductor-ferromagnet-superconductor (SFS) junction with the homogeneous magnetization of the central ferromagnetic layer. In contrast, it known to be  a fingerprint of the complicated spin structures, like ferromagnetic domains or multilayer in Josephson spin valves~\cite{krasnov,reverse,jsv1,jsv2,jsv3}. Fig.~\ref{3} also excludes any possibility for the unwanted shortings of the In leads, since a simple In-In junction can not demonstrate the observed $dV/dI(B)$ reversal.

$dV/dI(B)$ reversal  can also be demonstrated  by colormaps in Figs.~\ref{4} (a,b) and (d,e) for samples S1 and S2, respectively. The colormaps are obtained from $dV/dI(I)$ curves at fixed magnetic field values, which are  changed point-by-point in up or down directions. To establish definite sample magnetization state, every magnetic field sweep cycle begins from high field value $B$ = $\pm$100~mT. Due to the procedure, $dV/dI(B)$ reversal  is not connected with any time-dependent relaxation.  The panels (a) and (b) differ by the magnetic field sweep directions in Fig.~\ref{4}, which is from negative to positive values in (a) and is just opposite in (b). The previously described $dV/dI(B)$ reversal can be clearly seen, e.g., by the asymmetric black feature at $\pm$9~mT in Figs.~\ref{4} (a,b). The reversal effect is even more pronounced in (d) and (e) for normal magnetic fields.

\begin{figure}
\includegraphics[width=\columnwidth]{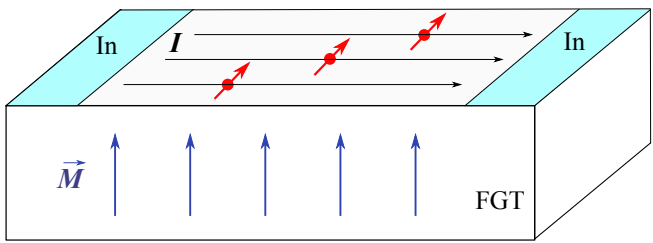}
\caption{(Color online) Sketch of the Josephson spin valve, which is realized in In-FGT-In junctions due to the spin-polarized surface state in the magnetic nodal-line topological semimetal FGT.  Supercurrent (partially) flows through the spin polarized surface state (grey region) with complicated spin polarization, while spin-dependent scattering with the magnetized FGT bulk is responsible for the spin valve behavior. }
\label{6}
\end{figure}

For Josephson effect, an important information can be obtained from the maximum supercurrent $I_c$ suppression. In principle, zero-resistance black region in the colormaps reflects the critical current $I_c(B)$  suppression pattern, as it is emphasized by the white envelope curves in  Figs.~\ref{4} (a,b) and (d,e). To obtain $I_c$ with high accuracy at fixed $B$, we sweep the current ten times from the zero value (i.e. from the superconducting $dV/dI$ = 0 state) to some value well above the $I_c$ (the resistive $dV/dI > 0$ state) and then determine $I_c$ as an average value of $dV/dI$ breakdown positions. The result is presented in Figs.~\ref{4} (c) and (f) for two magnetic field orientations, respectively.  The general $I_c(B)$ shape is asymmetric in both cases,  the asymmetry is reversed for the up (blue) and down (red) field sweeps. $I_c(B)$ also does not exhibit a conventional Fraunhofer pattern~\cite{aperiodicsfs1,aperiodicsfs2}.

There are also some features in Fig.~\ref{4}, which are different in two magnetic field orientations. 

For the in-plane magnetic fields, $I_c(B)$ shows fast aperiodic fluctuations in Fig.~\ref{4} (a-c). No distinct period could be detected at least for the field step as small as $\Delta B$ = 0.01~mT. We check, that our procedure gives $I_c$ values, which are perfectly stable at fixed magnetic field, as demonstrated in Fig.~\ref{5} (a). The maximum $I_c$ deviation over 1000 curves is about 0.005~mA at $B$ = -1.4~mT, which is negligible comparing to the observed fluctuations' amplitude 0.05~mA in Fig.~\ref{4} (c). Thus, the fluctuations are controlled by the external magnetic field, although they are found to be aperiodic. 

On the contrary, no noticeable fluctuations can be observed for normal magnetic field orientation, see Figs.~\ref{4}(d-f). The curves for the up (blue) and down (red) sweeps are reversed, but in addition there is an interplay between maximum and minimum in $I_c(B)$ at $\pm$12~mT, which is well known for the Josephson spin valves~\cite{krasnov,reverse,jsv1,jsv2,jsv3}.
 
Temperature dependence of the critical current $I_c(T)$ is shown in Fig.~\ref{5}(b). It closely reminds the temperature dependencies observed in a half-metallic CrO$_2$ based long Josephson junctions~\cite{cro2}, where a large contribution of spin triplet supercurrent was implied.

\section{Discussion}

As a result, we observe $I_c(B)$ pattern asymmetry and its' reversal in dependence on the magnetic field sweep direction. This effect can be observed for both magnetic field orientations, while in normal magnetic fields there is also a prominent change of the $I_c(B)$ shape during remagnetization. 

This behavior can not be expected for usual SFS junctions with the homogeneous magnetization of the central ferromagnetic layer, where remagnetization can only shift the $I_c(B)$ pattern position in magnetic field~\cite{cro2,reverse}.  On the other hand, the observed behavior is a known fingerprint of Josephson spin valves~\cite{krasnov,reverse,jsv1,jsv2,jsv3}. While in conventional Josephson junctions supercurrent is modulated by magnetic flux, in JSVs it is mainly defined by the relative orientation of magnetic layers, giving rise to the $I_c(B)$ asymmetry and reversal.

A conventional spin valve, in its simplest form, is a layered structure consisting of a thick (fixed) and a thin (free) ferromagnetic layers~\cite{valve1,valve2}. Spin valve resistance is defined by the relative angle between magnetizations of the layers due to the spin-dependent scattering, which can be tuned by field or flowing current. Spin valve can be naturally realized in  different types of topological materials and their heterostuctures with ferromagnets~\cite{timnal,wte,cosns,bite}. In this case, spin-polarized topological surface state acts as one layer of a spin valve, while the role of the other is played by the ferromagnetic lead or by the ferromagnetic sample's bulk~\cite{timnal,cosns}. The spin-polarized surface state acts as a source of spin, while spin-dependent scattering within the ferromagnet results in a different resistance depending on the magnetization direction~\cite{molenkamp2021}.

In the case of FGT, the presence of spin-polarized topological Fermi arcs has been demonstrated by ARPES~\cite{kim}, while spin momentum locking~\cite{spinmomlock} was inferred to be responsible for anti-symmetric magnetoresistance in FGT/graphite/FGT heterostructures~\cite{asymmr}. Thus, a FGT flake may be regarded as a spin valve, this scenario is independently verified by magnetoresistance of a single Au-FGT junction for the reference Hall bar sample in Fig.~\ref{5}(c), where typical spin-valve hysteresis is observed~\cite{valve1,valve2}. Moreover, Fig.~\ref{4} shows asymmetric resistive features even at high currents, i.e. for the suppressed superconductivity (for 2-10 Ohms junction resistance). These features are also reversed for two magnetic field directions, which confirms   spin-valve behavior in FGT. 

If a spin valve is sandwiched between two superconducting electrodes~\cite{krasnov,reverse,jsv1,jsv2,jsv3}, see Fig.~\ref{6}, asymmetric $I_c(B)$ pattern should be reversed after remagnetization. Effectiveness of singlet-triplet conversion depends on magnetic orientation misalignment, so $I_c(B)$ pattern depends on the spin-valve configuration. Due to the hysteresis in magnetization of a spin valve~\cite{valve1,valve2}, $I_c(B)$ demonstrates a mirror reversal in the opposite field sweeps~\cite{krasnov,reverse}.

The observed interplay between the $I_c(B)$ maximum and minimum after remagnetization in Fig.~\ref{4}(d-f) is very unusual. Generally, this behaviour requires breaking of certain symmetries. For FGT,  the inversion symmetry breaking is known at the interface~\cite{skyrm4}. This is supported by a number of experimental observations of the skyrmion-like spin textures, e.g., Bloch-type~\cite{skyrm1} and N\'eel-type~\cite{skyrm4,skyrm2,skyrm3} skyrmions, domain wall twists~\cite{skyrm5}, and chiral spin textures~\cite{skyrm6,skyrm7}. Inversion symmetry breaking in a system with a large spin-orbit interactions gives rise to the spin-orbit torque, comprising terms which are even and odd in magnetization. The relative signs of the terms changes under remagnetization, violating the reversal of the $I_c(B)$ pattern in Fig.~\ref{4}(d-f). This interplay in $I_c(B)$ is not observed in Fig.~\ref{4} (a,b,c), since the  FGT magnetization is not collinear to the out-of-plane current-induced polarization in this case.  We wish to note, that one can not ascribe the observed interplay to the spin valve memory effect~\cite{krasnov,jsv1}, since every remagnetization process starts from the same $B$ = $\pm$100~mT in our experiment.

Regarding the effects of the domain structure, one should note that the presence of several ferromagnetic domains
between the superconducting leads could generally give rise to essentially the same physics as in a 
JSV~\cite{dw1,dw2}. In particular, asymmetric non-Fraunhofer $I_c(B)$ patterns in SFS junctions with a complex 
multi-domain structure have been reported before~\cite{aperiodicsfs1,aperiodicsfs2}. However, the domain 
structure effects are hardly responsible for the results obtained in this paper. Sufficiently thick FGT samples
within the low-temperature range $\alt 5$K contain several types of domains \cite{domain1,domain2}, 
among which bubble-like domains with comparatively small sizes, about a few hundred nanometers, are randomly 
distributed over the surface, introducing a substantial stochastic component to the domain structure 
\cite{domain2}. We would like to emphasize here, that the asymmetric $dV/dI(B)$ curves and $dV/dI(B,I)$ 
colormaps in Figs.~\ref{3},~\ref{4} are highly reproducible, and therefore should not be attributed to any 
stochastic interfacial domain structures that would prevent to reproduce the results with an observed accuracy.
Although, the noncoplanar spin textures~\cite{PTHE} could noticeably contribute to aperiodic variations 
presented in Figs.~\ref{4} (a,b,c) for the in-plane field orientation.

Thus, our experimental results can be regarded as demonstration of the JSV, which is realized in the magnetic nodal-line topological semimetal FGT. Moreover, surface transport was ubiquitously attributed to carry Josephson current at long distances in JJs based on topological materials~\cite{topojj1,topojj2,topojj3,topojj4,topojj5}, which supports the overall interpretation.

\section{Conclusion}

As a conclusion, we investigate lateral electron transport between two 3~$\mu$m spaced superconducting In leads beneath a thick three-dimensional FGT exfoliated flake. At low 30~mK temperature, we observe Josephson supercurrent that exhibits unusual critical current $I_c$ suppression by the magnetic field $B$. The overall $I_c(B)$ pattern is asymmetric in respect to the B sign. We demonstrate, that the asymmetry is defined by the magnetic field sweep direction, so the $I_c(B)$ pattern is strictly reversed (as $B$ to $-B$ inversion) for the opposite sweeps. We also observe an interplay between maximum and minimum in $I_c(B)$ in normal magnetic fields, while there are fast aperiodic $I_c(B)$ fluctuations for the in-plane ones. These effects can not be expected for homogeneous superconductor-ferromagnet-superconductor junctions, while they are known for Josephson spin valves. The mostly possible scenario for Josephson spin valve realization in FGT is the misalignment of spin polarizations of the Fermi arc surface states and ferromagnetic FGT bulk, but we also discuss possible influence of spin-dependent transport between  magnetic domains.

\acknowledgments
We wish to thank V.T.~Dolgopolov and A.S.~Melnikov for fruitful discussions, and S.S~Khasanov for X-ray sample characterization.  We gratefully acknowledge financial support  by the  RF State task.

\end{document}